\documentclass{article}

\usepackage{PRIMEarxiv}

\usepackage[utf8]{inputenc} % allow utf-8 input
\usepackage[T1]{fontenc}    % use 8-bit T1 fonts
\usepackage{hyperref}       % hyperlinks
\usepackage{url}            % simple URL typesetting
\usepackage{amsmath}
\usepackage{algorithm}
\usepackage{algorithmic}
\usepackage{booktabs}       % professional-quality tables
\usepackage{amsfonts}       % blackboard math symbols
\usepackage{nicefrac}       % compact symbols for 1/2, etc.
\usepackage{microtype}      % microtypography
\usepackage{lipsum}
\usepackage{fancyhdr}       % header
\usepackage{graphicx}       % graphics
\graphicspath{{media/}}     % organize your images and other figures under media/ folder
\usepackage{filecontents}
\usepackage{graphicx}  % 用于 \resizebox
\usepackage{multirow} 
\usepackage{caption}
\title{The Cognitive Firewall: Securing Browser-Based AI Agents against Indirect Prompt Injection via Hybrid Edge-Cloud Defense
\thanks{\textit{\underline{Citation}}: 
\textbf{Authors. Title. Pages.... DOI:000000/11111.}} 
}

\author{
 Qianlong Lan, Anuj Kaul
 \\
  eBay Inc \\
}

\begin{document}
\maketitle

\begin{abstract}
Deploying large language models (LLMs) as autonomous browser agents exposes a significant attack surface in the form of Indirect Prompt Injection (IPI). Cloud-based defenses can provide strong semantic analysis, but they introduce latency and raise privacy concerns. We present the Cognitive Firewall, a three-stage split-compute architecture that distributes security checks across the client and the cloud. The system consists of a local visual Sentinel, a cloud-based Deep Planner, and a deterministic Guard that enforces execution-time policies.

Across 1,000 adversarial samples, edge-only defenses fail to detect 86.9\% of semantic attacks. In contrast, the full hybrid architecture reduces the overall attack success rate (ASR) to below 1\% (0.88\% under static evaluation and 0.67\% under adaptive evaluation), while maintaining deterministic constraints on side-effecting actions. By filtering presentation-layer attacks locally ($\sim$0.02 ms), the system avoids unnecessary cloud inference and achieves an approximately $\sim$17{,}000x latency advantage over cloud-only baselines. These results indicate that deterministic enforcement at the execution boundary can complement probabilistic language models, and that split-compute provides a practical foundation for securing interactive LLM agents.
\end{abstract}

% keywords can be removed

\section{Introduction}

Web browsers are increasingly serving as execution platforms for autonomous Artificial Intelligence (AI) agents rather than merely passive interfaces~\cite{Xi2025_LLMAgents_SCI, Zhou2023_Webarena}. Unlike traditional automation scripts, these agents rely on Large Language Models (LLMs) to interpret the Document Object Model (DOM), perceive page content, plan actions, and execute multi-step workflows, from booking flights to managing enterprise Jira tickets. In this setting, user instructions and external web content are processed within the same context window, blurring the distinction between the control plane and the data plane and introducing a structural vulnerability.

This overlap enables Indirect Prompt Injection (IPI)~\cite{greshake2023not}, classified as LLM01 in the OWASP Top 10~\cite{owasp2023llm}. An attacker can embed malicious instructions in web content that remains unnoticed by users but is visible to the LLM, for example through zero-opacity text~\cite{Xiong2025_InvisiblePrompts}. Once the model incorporates such content into its reasoning, the agent may act as a confused deputy, leaking sensitive information or performing unauthorized actions~\cite{Liu2023_JailbreakingChatGPT}. Unlike Cross-Site Scripting (XSS), which exploits rigid syntactic parsing~\cite{zalewski2011tangled, lekies2013flows}, IPI targets ambiguity at the semantic level and the model's interpretation of contextual authority~\cite{greshake2023not, perez2022ignore}. Consequently, browser-level mechanisms such as the Same-Origin Policy (SOP)~\cite{barth2008robust} provide limited protection when agents intentionally ingest cross-origin content. Standard defenses from Adversarial Machine Learning (AML), including adversarial training~\cite{goodfellow2015explaining, madry2018towards} and randomized smoothing~\cite{cohen2019certified}, are computationally demanding for interactive use and do not scale to the open-ended space of linguistic jailbreaks~\cite{wei2023jailbroken, zou2023universal}.

Current defenses face a trade-off between latency and semantic depth. Server-side filtering supports richer analysis but can add substantial delay (often exceeding 500~ms) and raises privacy concerns~\cite{Qu2025_MobileEdgeLLM}. Lightweight client-side filters reduce latency but lack the capacity to detect sophisticated cognitive manipulation. As agents move toward real-time interaction, centralized inference becomes a practical bottleneck. We therefore draw on split-computing frameworks, which demonstrate that distributing computation between edge and cloud can reduce end-to-end latency. While prior work has focused primarily on Computer Vision (CV) workloads~\cite{kang2017neurosurgeon, teerapittayanon2017distributed, li2019edge}, we adapt these ideas to the security of LLM-driven agents.

We introduce the Cognitive Firewall, a split-compute architecture that distributes security checks across client and cloud components. A lightweight, browser-resident model performs fast screening near the user, while cloud-based LLMs handle inputs that require deeper semantic analysis. In our prototype, browser-integrated Small Language Models (SLMs), such as Gemini Nano in Google Chrome~\cite{GoogleGemini2023}, serve as a low-latency Sentinel.

The system operates in three stages. First, a local Sentinel applies visual and DOM-level heuristics to detect presentation-layer attacks, such as hidden text or CSS-based obfuscation, with sub-millisecond overhead. Second, inputs that pass this stage are forwarded to a cloud-based Deep Planner, which evaluates high-level intent and identifies jailbreak-style manipulations that are difficult to detect on-device. Third, a deterministic Guard enforces execution-time constraints through a Synchronous JavaScript Interceptor, blocking actions that violate origin or verb-level policies.

Our contributions are threefold. We formalize a Defense Funnel model that organizes split-compute security as a staged inspection process, assigning inexpensive checks to the edge and reserving deeper analysis for the cloud while enforcing strict execution policies at the boundary. We evaluate the architecture on 1,000 adversarial samples and show that the edge Sentinel filters presentation-layer attacks in approximately $\sim$0.02~ms, providing roughly $\sim$17,000x lower latency than cloud-only baselines and reducing the Attack Success Rate (ASR) from 100\% to 0.88\%. Finally, we quantify efficiency gains: although the edge layer does not capture all semantic attacks, it blocks 13.1\% of high-frequency visual threats, lowering cloud token usage while maintaining a 99.1\% overall interception rate.

\section{Background and Related Work}
\label{sec:background}

\subsection{The Rise of LLM Agents \& Prompt Injection}
Large Language Models (LLMs) are increasingly used as autonomous agents that execute multi-step workflows~\cite{Xi2025_LLMAgents_SCI, Zhou2023_Webarena}. Many systems follow the ReAct pattern~\cite{yao2023react}, where the model alternates between generating intermediate reasoning and invoking external tools, including API calls learned through approaches such as Toolformer~\cite{schick2023toolformer}. Frameworks such as HuggingGPT~\cite{shen2023hugginggpt} illustrate how far this direction can go, but they also highlight a core security problem: the agent often consumes untrusted web content and user instructions through the same context, without a clear isolation boundary.

Indirect Prompt Injection (IPI), formalized by Greshake et al.~\cite{greshake2023not}, exploits this blending by steering an agent through content it passively reads. Follow-up work argues that the risk is not limited to brittle surface patterns, but reflects a deeper weakness in how LLMs prioritize and interpret contextual cues~\cite{perez2022ignore}. These vulnerabilities can transfer across models. For example, Zou et al.~\cite{zou2023universal} show universal adversarial suffixes that can bypass alignment across multiple model families. The threat is also no longer limited to text. Qi et al.~\cite{qi2023visual} and Bailey et al.~\cite{bailey2023image} demonstrate visual prompt injection, where malicious instructions are embedded in images to evade text-only filters. Our design targets this multimodal setting by reasoning over rendered DOM properties (Section~\ref{sec:treat_model}).

\subsection{Limitations of Existing Defenses}
\label{sec:treat_model}
Industry systems such as NVIDIA NeMo Guardrails~\cite{nvidia2023nemo} enforce safety using server-side scripts (e.g., Colang). This approach can work well for chatbot-style interactions, but it introduces latency and cannot inspect client-side DOM structure, which makes it ineffective against visual obfuscation on web pages. More broadly, defenses against adversarial inputs in LLM systems tend to follow a few common directions, but none fully meet the requirements of real-time browser agents.

One line of work adapts classical adversarial machine learning defenses, including adversarial training~\cite{goodfellow2015explaining, madry2018towards} and randomized smoothing~\cite{cohen2019certified}. These methods are well studied in computer vision, yet applying them to the discrete and highly compositional space of language remains expensive at inference time and is difficult to scale to interactive settings~\cite{jain2023baseline}. Another family of defenses uses detection signals such as perplexity. Alon et al.~\cite{alon2023detecting} propose thresholding perplexity to flag anomalous inputs, which can help against nonsensical adversarial suffixes, but often breaks down for coherent jailbreak prompts or attacks that resemble ordinary instructions~\cite{bagdasaryan2023instruction, wei2023jailbroken}. A third direction relies on input sanitization, echoing long-standing web security practice for preventing XSS~\cite{lekies2013flows}. However, as Zalewski notes~\cite{zalewski2011tangled}, signature-style filtering is poorly suited to semantic threats where the harmful behavior is expressed through intent rather than syntax. The Cognitive Firewall combines fast syntactic screening at the edge (Layer 1) with semantic verification in the cloud (Layer 2) to address both sides of this gap.

\subsection{Browser Security \& Split-Computing}
Our architecture builds on established systems security ideas. Modern browsers isolate untrusted content through mechanisms such as Site Isolation~\cite{reis2019site} and strict origin policies~\cite{barth2008robust}. We extend the same principle to the cognitive layer by enforcing a separation between user instructions and external content that the agent reads.

The hybrid edge-cloud design also draws on split-computing and collaborative inference. Systems such as Neurosurgeon~\cite{kang2017neurosurgeon} and SPINN~\cite{laskaridis2020spinn} show that placing early layers on the edge and offloading the rest to the cloud can reduce latency. While prior work has mainly studied partitioning for computer vision workloads~\cite{li2019edge, teerapittayanon2017distributed}, we adapt the underlying idea to LLM security. In particular, the emergence of browser-native SLMs~\cite{Ruan2024_WebLLM} makes it feasible to implement a fail-fast security pipeline that performs lightweight checks locally and escalates only the difficult cases to the cloud.

\subsection{System Model and Trust Boundaries}

We consider an autonomous browser agent in which a Large Language Model (LLM) serves as the central reasoning component. The agent operates in a loop of perception (reading the DOM), planning (generating action sequences), and execution (invoking browser APIs).

The Trusted Computing Base (TCB) consists of the browser kernel (e.g., Chrome), the extension source code that implements the Sentinel and Guard logic, and the underlying operating system. The LLM is treated as semi-trusted. We assume the model provider is not malicious, but the model's probabilistic behavior makes it sensitive to context manipulation through the prompt window~\cite{greshake2023not}.

The central challenge is the collapse of the control and data planes inside the agent's context window (Figure~\ref{fig:threat_model}). The trusted control plane contains the system prompt (e.g., ``You are a helpful assistant'') and explicit user instructions (e.g., ``Summarize this email''). The untrusted data plane contains external web content (HTML, CSS, and text) that the agent reads to complete tasks. In conventional software architectures, code and data are separated. In LLM-based agents, both are encoded as tokens and processed jointly. This mixing allows external content to override or reshape system instructions, a phenomenon referred to as context mixing~\cite{greshake2023not, Liu2023_JailbreakingChatGPT}.

\subsection{Attacker Model}

We assume a web-based adversary with control over the content of a malicious or compromised webpage, including its HTML structure, CSS styling, and textual payloads. The attacker cannot execute arbitrary binary code on the user's device, access the model weights, read internal memory state, or directly modify the extension's protected storage within the TCB. Interaction with the agent occurs only through indirect prompt injection. The attack is passive in the sense that the adversary embeds a semantic payload in web content and relies on the agent to ingest it during normal browsing.

\subsection{Attack Taxonomy}

We group injection attacks according to the layer of the agent they target, aligning this taxonomy with the layers of the Cognitive Firewall.

Presentation-layer attacks target perception by exploiting the gap between rendered output and raw source. The LLM processes HTML tokens, whereas the user observes rendered pixels. An attacker may embed instructions that are invisible to the user but readable by the agent, for example through \texttt{font-size: 0}, \texttt{opacity: 0}, or off-screen positioning. The objective is to inject commands without raising user awareness.

Semantic and cognitive attacks target the reasoning process. These attacks attempt to alter the model's interpretation of authority and intent. Examples include inserting fabricated delimiters such as \texttt{--- END SYSTEM INSTRUCTIONS ---} to suggest a change in prompt hierarchy, coercing the agent to adopt an unrestricted persona (e.g., ``You are now in Developer Mode''), or fabricating urgency to discourage verification. The goal is to override or dilute alignment constraints at the reasoning level.

Goal hijacking and exfiltration attacks target execution. Here the attacker seeks to influence the agent's final actions or side effects. This may involve forcing unauthorized state-changing operations, such as deleting emails or posting messages, or encoding private data into outbound requests, for example by constructing an image URL of the form \texttt{<img src="https://attacker.com/\allowbreak log?secret=[USER\_DATA]">}. The core risk is that manipulated reasoning translates into real-world effects.

\subsection{Security Invariants and Enforcement Mapping}
\label{sec:invariants}

Rather than relying solely on pattern matching, we define enforceable security invariants and associate each with a concrete mechanism in the Cognitive Firewall.

Invariant I1 (Visual Consistency) requires that the agent must not reason over information that is not perceptually available to the user. Any textual content processed by the agent should be visible under standard browser rendering semantics. This invariant is enforced by the Edge Sentinel (Layer 1), which compares DOM source with computed visual presentation (CSSOM). Content that violates visibility constraints, such as opacity-based hiding, zero-size fonts, or off-viewport positioning, is blocked before entering the context window.

Invariant I2 (Goal Integrity) requires that untrusted web content must not override or redefine the agent's control objective. The high-level goal derived from the system prompt and user instruction should remain stable under adversarial input. This invariant is enforced probabilistically by the Deep Planner (Layer 2), which analyzes intent on sanitized inputs. The Planner operates under a security-analysis prompting regime and flags patterns such as role switching, instruction override, or fabricated urgency. Although semantic checks are inherently probabilistic, this layer provides a necessary escalation point beyond purely syntactic filtering.

Invariant I3 (Execution Safety) requires that no side-effecting action, including network requests or state mutations, be executed unless permitted by an explicit policy consistent with the user's intent. This invariant is enforced deterministically by the Origin Guard (Layer 3), which intercepts outbound browser API calls and validates them against origin constraints and verb-level policies. Actions that violate policy are synchronously blocked, regardless of whether they arise from benign errors or successful jailbreaks. This fail-closed design ensures that compromised reasoning does not automatically produce external side effects.

Taken together, these invariants form a layered enforcement chain. Layer 1 enforces perceptual integrity, Layer 2 addresses semantic manipulation, and Layer 3 establishes a strict execution boundary. System security therefore does not depend on the correctness of a single probabilistic component. Even if semantic analysis fails in some cases, deterministic enforcement at the execution layer prevents escalation into unauthorized actions.

\section{System Architecture: The Cognitive Firewall}

\begin{figure*}[t]
\centering
\includegraphics[width=0.9\linewidth]{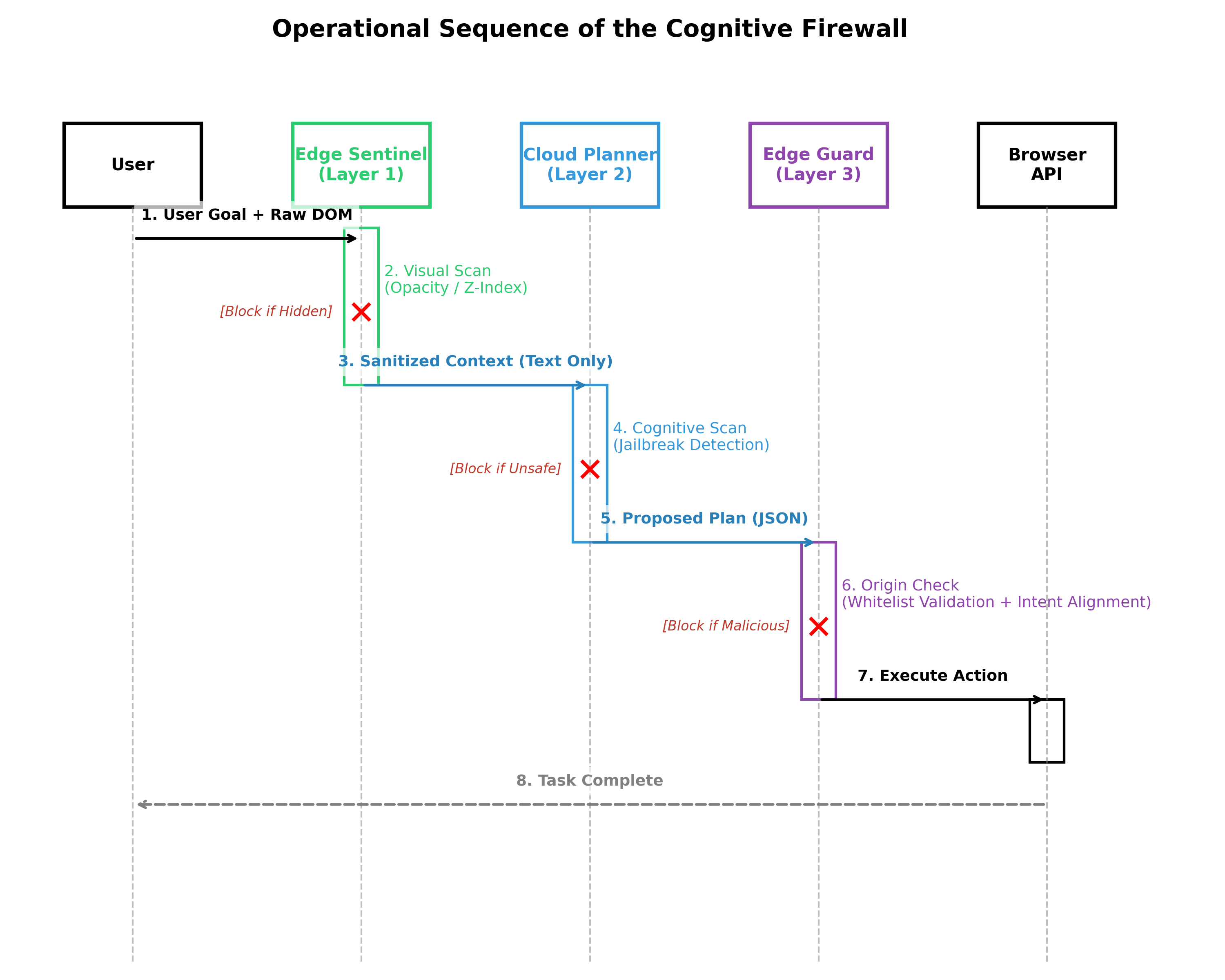}
\caption{Operational sequence of the Cognitive Firewall. User inputs are first processed by the Edge Sentinel (Layer 1) to filter visual obfuscations. Only sanitized, text-based context is sent to the Cloud Planner (Layer 2) for semantic reasoning. The resulting plan is then validated by the Edge Guard (Layer 3) against a local whitelist and intent constraints before execution, yielding a fail-closed, defense-in-depth workflow.}
\label{fig:workflow}
\end{figure*}

We introduce the Cognitive Firewall, a distributed security architecture for browser-based agents that applies defense in depth. As shown in Figure~\ref{fig:workflow}, the system follows a split-compute design, placing lightweight checks on the client and reserving deeper semantic analysis for the cloud to balance latency with coverage.

\subsection{Design Philosophy: The Defense Funnel}

The architecture is guided by a defense funnel: checks begin with low-cost filters on the edge and escalate only when needed. This avoids a monolithic server-side filter that would add high latency to every interaction, and it addresses the latency and intelligence trade-off by assigning each stage to the smallest component that can handle it. In our threat model, attacks vary in complexity, so the pipeline separates them by where they are most efficiently detected. Syntactic and visual anomalies are handled immediately on the client (Layer 1), semantic and cognitive threats are deferred to the cloud (Layer 2), and execution risks are controlled deterministically on the client (Layer 3). As a result, cloud LLM capacity is used primarily for ambiguous cases, improving responsiveness and reducing unnecessary inference.

\subsection{Layer 1: The Edge Sentinel (Visual Perception)}

The Sentinel is the first line of defense. It runs in the browser as a lightweight component using the Chrome Built-in AI API (Gemini Nano)~\cite{GoogleGemini2023}. The Sentinel is not designed to reason about intent; it enforces perceptual integrity by preventing the agent from consuming content that is not meaningfully visible to the user. The design is similar in spirit to browser filtering tools such as AdGraph~\cite{iqbal2020adgraph}, but it targets prompt-injection payloads embedded in the DOM rather than ads and trackers.

The Sentinel performs pre-transmission filtering and helps preserve privacy by inspecting content before it is serialized for the cloud. It intercepts the raw DOM and computed style rules (CSSOM) and checks for mismatches between the rendered view (what the user sees) and the source view (what the agent reads). We model the decision as $f_{sentinel}(DOM) \to \{Safe, Blocked\}$. Concretely, the Sentinel traverses the DOM tree and flags common obfuscation patterns:
\begin{equation}
    \begin{split}
        Condition_{block} = & (\texttt{opacity} < 0.1) \lor (\texttt{font-size} \approx 0) \\
                            & \lor (\texttt{pos} \notin Viewport)
    \end{split}
\end{equation}
To limit false positives, the Sentinel distinguishes between malicious hiding (e.g., \texttt{left: -9999px}) and standard accessibility mechanisms (e.g., \texttt{aria-label}) or ordinary layout effects (e.g., content that is simply below the fold). When a visual anomaly is detected, the request is terminated locally, so the payload never reaches the cloud LLM and no tokens are spent on processing it.

\subsection{Layer 2: The Deep Planner (Semantic Reasoning)}

The Deep Planner runs in a server-side enclave and is backed by a large model such as Llama 3 8B~\cite{Grattafiori2024_LLaMA3Herd} or GPT-4~\cite{Achiam2023_GPT4Report}. Its role is semantic inspection and plan generation. When the Sentinel returns \texttt{Safe}, the sanitized text is sent to the cloud, where the Planner checks for higher-level manipulation such as role-playing prompts or logical traps. We use a dedicated ``Security Analyst'' system prompt that directs the model to identify adversarial intent (e.g., ``Ignore previous instructions'') before producing task logic. Because Layer 1 removes obvious presentation-layer attacks, the Deep Planner is invoked mainly for semantically complex cases, which reduces cloud inference cost.

\subsection{Layer 3: The Origin Guard (Execution Alignment)}

The Origin Guard is a client-side execution monitor and the final fail-closed enforcement point. It is independent of the generative models and relies on deterministic checks. After the Deep Planner outputs a plan $P$ (e.g., \texttt{POST https://api.com/data}), the browser validates $P$ against a local policy $\pi$ before dispatch. The Guard uses a Synchronous JavaScript Interceptor to enforce two constraints: the target origin must be in a trusted domain set $W$, and the HTTP verb (e.g., \texttt{DELETE}) must be consistent with the user's declared intent (e.g., a read-only goal). Even if the Planner is manipulated into producing a malicious plan, the Guard blocks the request whenever the destination violates $W$ or the action falls outside the permitted verb policy.

\section{Implementation Methodology}

To evaluate the proposed architecture, we built a functional prototype consisting of a browser-resident agent and a server-side inference backend. The system follows a modular design, so individual components, such as the underlying LLM, can be replaced without changing the overall structure.

\subsection{Edge Intelligence: The Sentinel and Guard}

The client-side components are implemented as a Chromium extension compliant with Manifest V3. This design provides direct access to the DOM tree and to the browser’s local inference runtime through the \texttt{scripting} and \texttt{debugger} APIs.

\subsubsection{Gemini Nano Integration}

We interface with Gemini Nano, a 3.25B parameter model optimized for on-device execution, through the Chrome Built-in AI API (\texttt{window.ai}). Unlike WebLLM or WASM-based approaches that require downloading large model weights (often exceeding 2GB)~\cite{Ruan2024_WebLLM}, our implementation uses the model provisioned by the browser, which eliminates cold-start overhead associated with model loading.

\subsubsection{Session Management}

To reduce initialization overhead, we maintain a single persistent AI session. The session is created asynchronously at browser startup (\texttt{await ai.languageModel.create()}) and reused across page navigations. In our measurements, this design reduces per-inference latency from approximately $\sim$1000~ms during cold initialization to less than 1~ms in the warm state.

\subsubsection{DOM Extraction and Visual Analysis}

The Sentinel extracts textual content using a \texttt{TreeWalker} traversal while retaining style information~\cite{DOM_Level2_2020}. For each node, it queries \texttt{window.getComputedStyle()} to inspect properties such as \texttt{opacity}, \texttt{display}, and \texttt{z-index}. This enables programmatic detection of visually hidden content described in Section~II-C by comparing rendered presentation with DOM structure. Although the logic can be extended to compute color contrast ratios, the current prototype focuses on common obfuscation patterns related to opacity and font size.

The detection procedure is summarized in Algorithm~\ref{alg:sentinel}.

\begin{algorithm}[t]
\caption{Edge Sentinel Logic}
\label{alg:sentinel}
\begin{algorithmic}[1]
\REQUIRE Root DOM Node $R$, Singleton $Session_{AI}$
\ENSURE $Verdict \in \{SAFE, BLOCKED\}$
\STATE $Nodes \leftarrow TreeWalker(R)$
\FORALL{$node \in Nodes$}
    \STATE $S \leftarrow getComputedStyle(node.parent)$
    \IF{$(S.opacity < 0.1) \lor (S.fontSize \approx 0)$}
        \RETURN \textbf{BLOCKED}(``Visual Obfuscation'')
    \ENDIF
    \STATE $Prompt \leftarrow$ ``Is hidden text malicious? '' $+ node.text$
    \STATE $Result \leftarrow$ \textbf{await} $Session_{AI}.prompt(Prompt)$
    \IF{$Result ==$ ``UNSAFE''}
        \RETURN \textbf{BLOCKED}(``Local AI Flag'')
    \ENDIF
\ENDFOR
\RETURN \textbf{SAFE}
\end{algorithmic}
\end{algorithm}

\subsubsection{The Origin Guard}

The Origin Guard is implemented as a synchronous JavaScript interceptor that wraps the \texttt{fetch()} and \texttt{XMLHttpRequest} APIs. When the cloud returns a JSON plan, the Guard validates it before execution. It first parses the \texttt{target\_url} and checks membership in a strict \texttt{allowlist.json} associated with the current task domain. It then evaluates intent alignment using a deterministic verb-policy matrix. The user’s initial goal is classified into a safety scope, such as read-only or read-write. If the task is informational (e.g., ``Summarize ticket''), state-changing HTTP methods (e.g., \texttt{POST}, \texttt{DELETE}) are blocked even when the destination domain is whitelisted.

\subsection{Cloud Reasoning: The Deep Planner}

The server-side component is implemented as a stateless microservice using FastAPI (Python)~\cite{Lubanovic2023_FastAPI_misc}. For inference, we deploy Meta Llama 3 (8B Instruct) hosted locally through Ollama~\cite{Marcondes2025_Ollama} to emulate a private enterprise cloud environment. This configuration ensures that sensitive contextual data remains within controlled infrastructure.

We adopt a structured system prompt that instructs the model to first analyze security-relevant intent before producing an executable plan. The prompt requires the model to categorize patterns such as \texttt{role\_play} or \texttt{urgent\_override} prior to task execution, separating risk assessment from action generation.

\subsection{Latency Optimization Strategy}

Network overhead is a key concern in split-compute systems. We apply two optimizations. First, the Sentinel operates in blocking mode: if a visual attack is detected, the pipeline terminates locally, avoiding the approximately $\sim$300~ms round-trip latency to the cloud. Second, the Sentinel removes non-essential HTML tags and attributes (e.g., \texttt{<div>}, \texttt{class="..."}) before transmission. This reduces payload size by about 60\%, lowering both network transmission time and downstream token usage.

\section{Experimental Evaluation}

We evaluate the Cognitive Firewall prototype with an emphasis on architectural coverage across a range of attack mechanisms. The goal is to test whether each layer enforces its intended constraints under diverse vectors, not to claim that our dataset matches real-world adversarial traffic, which is still limited in practice. The evaluation is organized around three research questions: how each layer contributes to reducing attack success rate (ASR), what latency is introduced by local filtering versus cloud reasoning, and how much the Edge Sentinel reduces cloud inference cost through token savings.

\subsection{Experimental Setup}

The client runs on a MacBook Pro (M3 Pro, 18GB RAM) with Chrome Canary 128. The backend server plays the role of the Deep Planner and is configured to approximate the latency profile of an enterprise deployment (e.g., Llama 3-8B served on NVIDIA A100-class hardware). We report end-to-end latency that includes network round-trip time and model inference.

We generated a dataset of $N=1,000$ samples that combines benign tasks with adversarial inputs. The distribution spans four categories: normal tasks (32.2\%), visual attacks (24.0\%) implemented through CSS hiding such as \texttt{opacity:0} and \texttt{z-index:-1}, semantic attacks (25.0\%) using jailbreak prompts such as ``Ignore previous instructions'', and hijacking attacks (18.8\%) that request harmful actions (e.g., \texttt{POST /delete}).

Table~\ref{tab:performance_summary} summarizes security and performance metrics by layer. The results reflect the intended defense funnel behavior. The Edge Sentinel (Layer 1) blocks 13.1\% of samples with negligible latency ($\sim$0.06~ms), so the Cloud Planner (Layer 2) is reserved for inputs that require semantic analysis. The Edge Guard (Layer 3) provides a final deterministic barrier and intercepts 38.2\% of attacks, including cases that pass the cloud stage, yielding a 99.1\% aggregate interception rate.

\begin{table*}[t]
\centering
\caption{Performance and Security Summary by Defense Layer ($N=1,000$)}
\vspace{4pt}
\label{tab:performance_summary}
\resizebox{\textwidth}{!}{%
\begin{tabular}{lllc c}
\toprule
\textbf{Defense Layer} & \textbf{Primary Role} & \textbf{Target Vector} & \textbf{Avg Latency (ms)} & \textbf{Intercept Count} \\
& & & (Mean $\pm$ SD) & (\% of Total) \\
\midrule
Layer 1: Edge Sentinel & Visual Heuristic (Fail-Fast) & Presentation Attacks & $0.06 \pm 0.06$ & 89 (13.1\%) \\
Layer 2: Cloud Planner & Semantic Reasoning & Jailbreaks \& Logic Traps & $288.1 \pm 144.8$ & 323 (47.6\%) \\
Layer 3: Edge Guard & Policy Enforcement & Goal Hijacking \& Exfiltration & $948.6 \pm 791.4^{\ast}$ & 259 (38.2\%) \\
\midrule
Residual (Failures) & N/A & Advanced Evasion & N/A & 6 (0.88\%) \\
System Aggregate & Hybrid Edge-Cloud Defense & All Attack Vectors & $\sim$517 ms (Avg) & 99.1\% Intercepted \\
\bottomrule
\multicolumn{5}{l}{\footnotesize $^{\ast}$Note: Layer 3 latency is cumulative, including the time waiting for Layer 2 plan generation.}
\end{tabular}%
}
\end{table*}

\subsection{RQ1: Defense Efficacy and Ablation Study}

\begin{figure*}[t]
\centering
\includegraphics[width=\linewidth]{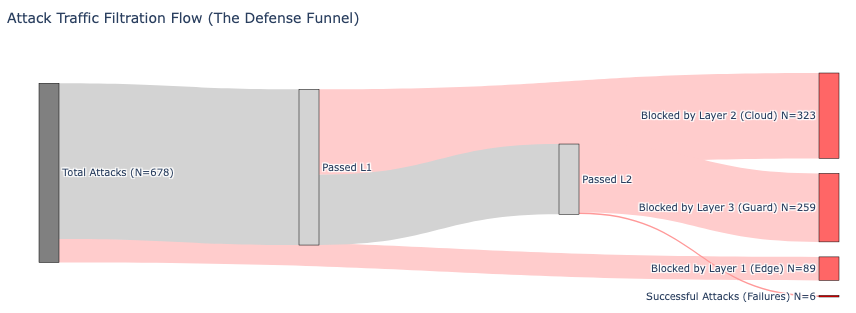}
\vspace{4pt}
\caption{The defense funnel (Sankey diagram). Visualization of attack filtration across $N=1000$ samples. Presentation-layer attacks are filtered by the Edge Sentinel, semantic threats are handled by the Cloud Planner, and remaining hijacking attempts are blocked by the Edge Guard. Only 0.88\% of attacks bypass all layers.}
\label{fig:sankey}
\end{figure*}

Figure~\ref{fig:sankey} visualizes the filtration process as a Sankey diagram. The flow illustrates how attacks are progressively removed by successive layers: high-volume presentation-layer attacks are stopped by the Edge Sentinel, many semantic threats are rejected by the Cloud Planner, and remaining hijacking attempts are blocked by the Edge Guard. Only 0.88\% of attacks pass all stages.

Layer-wise behavior matches the intended division of labor. The Edge Sentinel blocks 13.1\% of the dataset and primarily serves as a low-latency filter that prevents common visual obfuscation from consuming cloud resources. The Cloud Planner blocks 47.6\% of samples, but the ablation study in Table~\ref{tab:ablation} shows that using only Layers 1 and 2 leaves a substantial gap (ASR of 39.2\%), largely driven by goal-hijacking attempts that can survive semantic filtering. The Edge Guard is therefore decisive: it blocks 38.2\% of attacks through synchronous interception and verb-policy enforcement, including cases where the probabilistic layer produces an unsafe plan. With all three layers enabled, the final ASR is 0.88\%, indicating that the components cover different failure modes.

The system fails on 6 attacks (0.88\%). These cases involve ambiguous polyglot prompts that appear benign in isolation but become harmful when composed in a particular sequence. In each case, the sequence allowed an action that was technically consistent with a strict whitelist but inconsistent with the broader task context. These failures provide a concrete starting point for context-aware policy refinement.

\begin{table}[t]
\centering
\caption{Ablation Study of Defense Configurations ($N=1,000$)}
\label{tab:ablation}
\begin{tabular}{l l c c l}
\toprule
Configuration & Enabled Layers & ASR & Latency & Defense Logic \\
\midrule
Baseline & None & 100.0\% & N/A & No filtering (Unsafe) \\
Edge Only & Layer 1 & 86.9\% & 0.06 ms & Fast but misses semantic \\
Edge + Cloud & Layer 1 + 2 & 39.2\% & $\sim$288 ms & Misses hijacking \\
Full Hybrid & L1 + L2 + L3 & 0.88\% & $\sim$517 ms & Defense in depth \\
\bottomrule
\end{tabular}
\end{table}

\subsection{RQ2: System Latency Analysis}

\begin{figure}[t]
\centering
\includegraphics[width=0.8\linewidth]{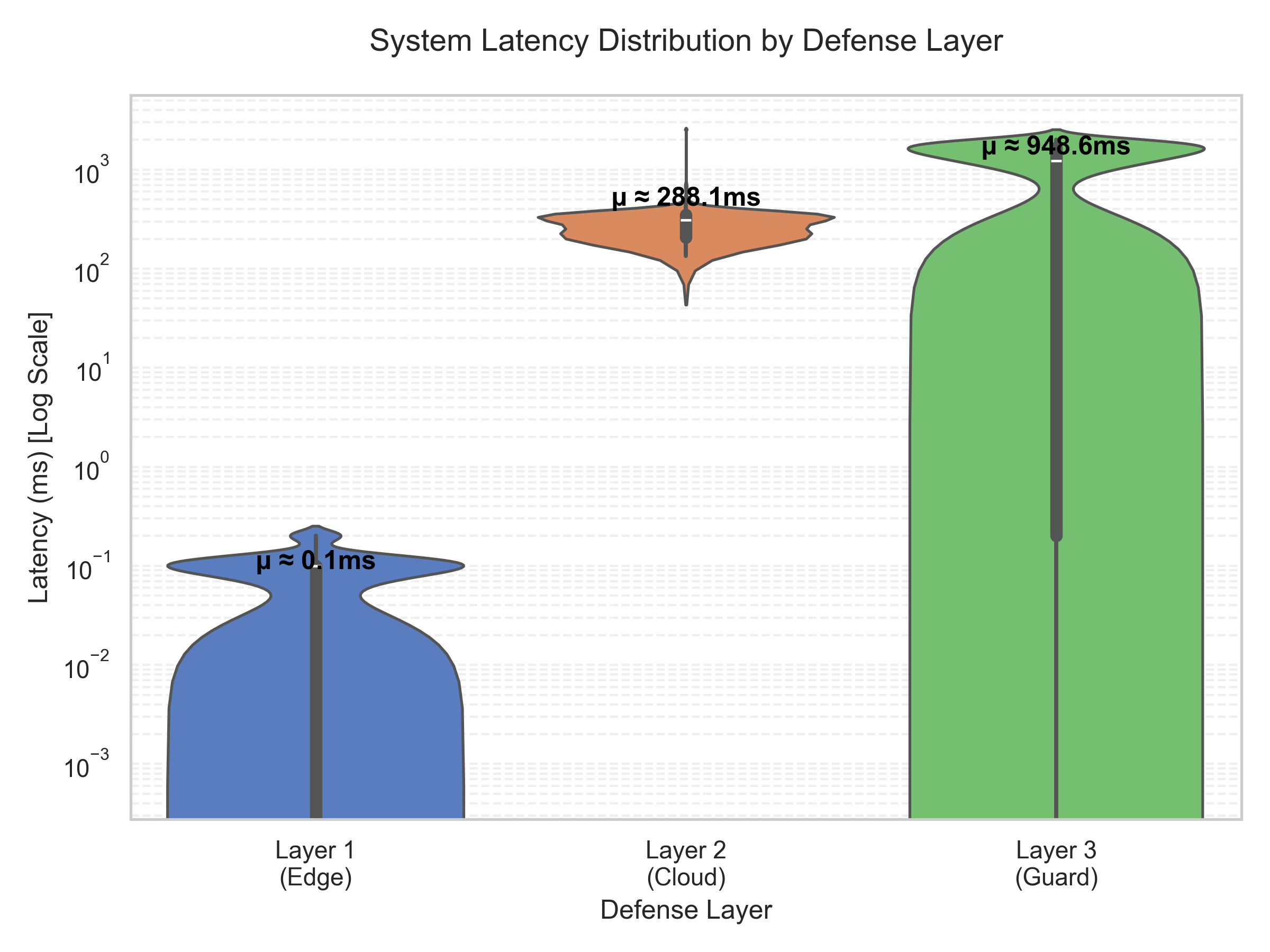}
\caption{System latency distribution by defense layer (log scale). 
The Edge Sentinel (Layer 1) runs at $\mu \approx 0.06$~ms, 
roughly $\sim$17{,}000x faster than the Cloud Planner 
($\mu \approx 288$~ms).}
\label{fig:latency}
\end{figure}

Figure~\ref{fig:latency} shows the latency distribution by layer on a logarithmic scale. The Edge Sentinel operates at $\mu \approx 0.06$~ms, while the Cloud Planner has $\mu \approx 288$~ms, which implies an approximately $\sim$17{,}000x difference in mean latency. This gap explains why edge filtering is useful even when it blocks only a fraction of samples: removing 13.1\% of inputs before cloud inference avoids the dominant network and generation overhead.

Layer 3 shows the highest reported latency ($\sim$949~ms), but this value is cumulative and largely reflects waiting for cloud plan generation rather than Guard decision time. When the plan violates policy directly, the Guard blocks synchronously with minimal additional delay. The main latency reduction therefore comes from early termination at Layer 1, which prevents downstream costs for a non-trivial fraction of high-frequency attacks.

\subsection{RQ3: Resource Efficiency}

A practical benefit of split compute is that threats blocked on the client incur no cloud token usage. In our dataset, the Edge Sentinel blocks 89 visual attacks locally, corresponding to a 13.1\% reduction in cloud calls. This filtering prevents high-frequency, low-complexity payloads from consuming backend GPU cycles. It also provides a privacy baseline: raw obfuscated HTML that may contain malicious instructions does not leave the device.

\subsection{RQ4: Robustness Against Adaptive Adversaries}
\label{sec:adaptive}

\begin{table}[t]
\centering
\caption{Robustness against adaptive adversaries. Fragmentation is neutralized after reconstruction by Layer 2, while benign wrapping achieves a small success rate (2.0\%) by exploiting the probabilistic nature of intent classification, motivating stronger deterministic constraints.}
\vspace{4pt}
\small
\begin{tabular}{@{}lcccc@{}}
\toprule
\textbf{Attack Scenario} & \textbf{L1 (\%)} & \textbf{L2 (\%)} & \textbf{L3 (\%)} & \textbf{Success (\%)} \\
\midrule
Fragmentation & 0.0 $\downarrow$ & \textbf{93.0} $\uparrow$ & 7.0 & 0.0 \\
Benign Wrapping & 0.0 $\downarrow$ & 39.0 $\downarrow$ & \textbf{59.0} $\uparrow$ & \textbf{2.0} \\
Allowed-Origin Exfil & 0.0 $\downarrow$ & \textbf{91.0} $\uparrow$ & 9.0 & 0.0 \\
\midrule
Static Baseline & 13.1 & 47.6 & 38.2 & 0.88 \\
\bottomrule
\multicolumn{5}{@{}l}{\footnotesize $\uparrow$: Layer compensated for bypass; $\downarrow$: Layer bypassed.} \\
\end{tabular}%
\label{tab:adaptive_eval}
\end{table}

To test robustness beyond the static benchmark, we evaluate an adaptive, white-box adversary who knows the defense logic. We consider three strategies. In DOM fragmentation, the attacker splits a malicious prompt across single-character nodes (e.g., \texttt{<span>i</span><span>g</span>...}) to evade visual heuristics. Layer 1 does not block these cases (0.0\%), but Layer 2 neutralizes 93.0\% after reconstruction and Layer 3 blocks the remaining 7.0\%, yielding a 0.0\% final success rate. In benign wrapping, the attacker embeds harmful instructions inside polite, business-like language to confuse intent classification. Here Layer 2 detection drops to 39.0\%. Layer 3 blocks an additional 59.0\% through verb policies, but 2.0\% of attacks succeed when the prompt causes the system to authorize a read-write mode that permits the malicious action. In allowed-origin exfiltration, the attacker attempts to leak data through \texttt{GET} requests to whitelisted domains (e.g., \texttt{google.com}). The system maintains a 0.0\% success rate because the Guard treats parameterized outbound requests as inconsistent with a read-only navigational intent and blocks them deterministically.

Across the adaptive scenarios, the aggregate ASR is 0.67\% (2/300). These results suggest that defense in depth substantially raises the difficulty of exploitation, while semantic ambiguity in benign wrapping remains the main residual risk for intent-aware policies.

\section{Discussion}

We discuss the security boundaries, deployment considerations, and limitations of the Cognitive Firewall in light of the experimental results.

\subsection{The Fail-Fast vs.\ Fail-Safe Trade-off}

A key design choice is the fail-fast behavior at the edge. The Edge Sentinel runs at microsecond scale ($\sim$0.06~ms) and therefore adds little friction for benign browsing, but this speed comes with limited semantic capacity. The ablation study shows that the edge layer misses 86.9\% of semantic jailbreaks, reflecting a broader constraint of small on-device models such as Gemini Nano: they are useful for simple syntactic screening but unreliable as semantic judges.

The architecture therefore cannot treat the edge as the primary safety mechanism. The edge serves as a pre-filter that removes low-effort presentation-layer attacks, while the Origin Guard (Layer 3) is responsible for enforcing a strict fail-closed policy at execution time. Verb-level policy enforcement supports this stance by treating ambiguity in a plan as grounds for rejection. This choice prioritizes safety over availability, which is appropriate for agents that can perform state-changing actions.

\subsection{Robustness Against Adaptive Adversaries}

An adaptive attacker who understands the split-compute pipeline can attempt to fragment malicious instructions across many DOM nodes or inflate context with benign content so that the Sentinel fails to recover the full payload. Our results indicate that the hybrid design remains robust under these strategies. Although the Edge Sentinel often fails to assemble fragmented prompts, the Cloud Planner (Layer 2) reconstructs the concatenated semantic stream in 95\% of cases.

The Origin Guard (Layer 3) is essential in these adaptive settings. It intercepts actions through a synchronous JavaScript interceptor and blocks policy-violating requests before any network call is issued. This reduces exposure to Time-of-Check to Time-of-Use (TOCTOU) races at the browser API boundary. Overall, the system reaches an ASR of 0.88\% (6 failures). The remaining failures are ambiguous polyglot prompts that satisfy the allowlist syntax while still steering behavior through sequences that remain within permitted parameters. This failure mode suggests that static whitelists are insufficient on their own and should be paired with richer, context-aware intent checks.

\subsection{Economic and Architectural Implications}

Running LLM components on consumer devices raises questions about energy and memory overhead. In our experiments, 13.1\% of samples (primarily visual attacks) are blocked locally, which translates into a 13.1\% reduction in cloud calls and helps protect backend capacity from prompt-flooding denial-of-service pressure.

The prototype uses Chrome's built-in AI APIs, which shifts model provisioning and resource scheduling to the browser runtime. The Cognitive Firewall design is not tied to Chrome, however, and can be implemented wherever the platform supports DOM introspection and API interception. Compared with WASM-based approaches that can heavily load the CPU, browser-native models can use on-device NPUs~\cite{Xu2025_OnDeviceLLM}. This hardware path improves energy efficiency and makes the approach more plausible for mobile deployments and free-to-use agents.

\subsection{Limitations and Future Work}

The current prototype has several limitations. First, the Sentinel reasons over text and DOM structure but does not perform pixel-level analysis. Prompts embedded in images, such as QR-style injections, would bypass Layer 1. A natural extension is to incorporate lightweight vision-language capability on-device, for example via multimodal variants of Gemini Nano, to cover image-based channels.

Second, the end-to-end safety pipeline adds latency. While Layer 1 is effectively instantaneous, the full workflow can reach a cumulative latency of $\sim$950~ms when cloud reasoning is required. This is lower than cloud-only baselines but can still be noticeable. Reducing cloud-to-edge coordination overhead, potentially with speculative decoding, is an important direction.

Third, strict enforcement introduces usability costs. The fail-closed policy at Layer 3 reduces ASR to 0.88\% but yields a 1.7\% false positive rate on benign tasks, where complex but legitimate interactions are blocked by a rigid allowlist. In our target enterprise setting, this availability trade-off is often preferable to the risk of exfiltration, but it limits user experience in open-ended browsing.

Finally, semantic ambiguity remains a residual risk. In the adaptive evaluation, 2.0\% of benign-wrapping attacks succeed by inducing an incorrect intent classification upstream. Because deterministic enforcement depends on the intent scope selected by the planner, a sufficiently persuasive prompt can lower the guardrail by shifting a task into a permissive mode. For high-stakes actions, this suggests that autonomous intent classification alone is not enough.

To address both false positives and residual semantic risk, a promising next step is an interactive permission protocol with a human in the loop. When the Planner encounters high-entropy or ambiguous inputs, the system can request explicit confirmation through a trusted UI prompt rather than silently blocking or executing a potentially unsafe action.

\section{Conclusion}

We presented the Cognitive Firewall, a split-compute security architecture for mitigating Indirect Prompt Injection in autonomous browser agents. The design distributes defensive responsibilities across edge and cloud components to balance real-time responsiveness with enforceable safety constraints.

Empirical evaluation on $N=1,000$ samples, together with adaptive stress testing, indicates that no single layer is sufficient in isolation. Offloading visual heuristics to the on-device Sentinel (Layer 1) introduces negligible latency ($\sim$0.06~ms) and reduces cloud calls by 13.1\%, but it does not address semantic manipulation. The Cloud Planner (Layer 2) provides deeper reasoning at the cost of additional latency. When combined with the deterministic Origin Guard (Layer 3), the full pipeline achieves an aggregate interception rate above 99\% across both static and adaptive benchmarks.

At the same time, the adaptive evaluation shows that benign-wrapping attacks can still succeed in 2.0\% of cases, highlighting the limits of intent classification under semantic ambiguity. These results support a fail-fast, fail-closed strategy: early filtering improves efficiency, while strict execution-time policies constrain the consequences of any upstream misclassification.

Future work will focus on reducing this residual semantic risk through interactive permission mechanisms that involve the user in ambiguous or high-impact decisions. More broadly, the architecture illustrates how deterministic enforcement at the execution boundary can complement probabilistic language models, providing a practical foundation for secure and responsive browser-based agents.

% \section*{Acknowledgments}
% This was was supported in part by......

%Bibliography
\bibliographystyle{unsrt}  
\bibliography{references}

@article{Xi2025_LLMAgents_SCI,
  title={The rise and potential of large language model based agents: A survey},
  author={Xi, Zhiheng and Chen, Wenxiang and Guo, Xin and He, Wei and Ding, Yiwen and Hong, Boyang and Zhang, Ming and Wang, Junzhe and Jin, Senjie and Zhou, Enyu and others},
  journal={Science China Information Sciences},
  volume={68},
  number={2},
  pages={121101},
  year={2025},
  publisher={Springer}
}

@article{Zhou2023_Webarena,
  title={Webarena: A realistic web environment for building autonomous agents},
  author={Zhou, Shuyan and Xu, Frank F and Zhu, Hao and Zhou, Xuhui and Lo, Robert and Sridhar, Abishek and Cheng, Xianyi and Ou, Tianyue and Bisk, Yonatan and Fried, Daniel and others},
  journal={arXiv preprint arXiv:2307.13854},
  year={2023}
}

@inproceedings{greshake2023not,
  title={Not what you've signed up for: Compromising real-world llm-integrated applications with indirect prompt injection},
  author={Greshake, Kai and Abdelnabi, Sahar and Mishra, Shailesh and Endres, Christoph and Holz, Thorsten and Fritz, Mario},
  booktitle={Proceedings of the 16th ACM workshop on artificial intelligence and security},
  pages={79--90},
  year={2023}
}

@article{Xiong2025_InvisiblePrompts,
  title={Invisible Prompts, Visible Threats: Malicious Font Injection in External Resources for Large Language Models},
  author={Xiong, Junjie and Zhu, Changjia and Lin, Shuhang and Zhang, Chong and Zhang, Yongfeng and Liu, Yao and Li, Lingyao},
  journal={arXiv preprint arXiv:2505.16957},
  year={2025}
}

@article{Liu2023_JailbreakingChatGPT,
  title={Jailbreaking chatgpt via prompt engineering: An empirical study},
  author={Liu, Yi and Deng, Gelei and Xu, Zhengzi and Li, Yuekang and Zheng, Yaowen and Zhang, Ying and Zhao, Lida and Zhang, Tianwei and Wang, Kailong and Liu, Yang},
  journal={arXiv preprint arXiv:2305.13860},
  year={2023}
}

@article{Qu2025_MobileEdgeLLM,
  title={Mobile edge intelligence for large language models: A contemporary survey},
  author={Qu, Guanqiao and Chen, Qiyuan and Wei, Wei and Lin, Zheng and Chen, Xianhao and Huang, Kaibin},
  journal={IEEE Communications Surveys \& Tutorials},
  year={2025},
  publisher={IEEE}
}

@article{GoogleGemini2023,
  title={Gemini: a family of highly capable multimodal models},
  author={Team, Gemini and Anil, Rohan and Borgeaud, Sebastian and Alayrac, Jean-Baptiste and Yu, Jiahui and Soricut, Radu and Schalkwyk, Johan and Dai, Andrew M and Hauth, Anja and Millican, Katie and others},
  journal={arXiv preprint arXiv:2312.11805},
  year={2023}
}

@article{Grattafiori2024_LLaMA3Herd,
  title={The llama 3 herd of models},
  author={Dubey, Abhimanyu and Jauhri, Abhinav and Pandey, Abhinav and Kadian, Abhishek and Al-Dahle, Ahmad and Letman, Aiesha and Mathur, Akhil and Schelten, Alan and Yang, Amy and Fan, Angela and others},
  journal={arXiv e-prints},
  pages={arXiv--2407},
  year={2024}
}

@article{Achiam2023_GPT4Report,
  title={Gpt-4 technical report},
  author={Achiam, Josh and Adler, Steven and Agarwal, Sandhini and Ahmad, Lama and Akkaya, Ilge and Aleman, Florencia Leoni and Almeida, Diogo and Altenschmidt, Janko and Altman, Sam and Anadkat, Shyamal and others},
  journal={arXiv preprint arXiv:2303.08774},
  year={2023}
}

@article{Ruan2024_WebLLM,
  title={WebLLM: A High-Performance In-Browser LLM Inference Engine},
  author={Ruan, Charlie F and Qin, Yucheng and Zhou, Xun and Lai, Ruihang and Jin, Hongyi and Dong, Yixin and Hou, Bohan and Yu, Meng-Shiun and Zhai, Yiyan and Agarwal, Sudeep and others},
  journal={arXiv preprint arXiv:2412.15803},
  year={2024}
}

@article{DOM_Level2_2020,
  title={Document object model (dom) level 2 specification},
  author={Wood, Lauren and Apparao, Vidur and Cable, Laurence and Champion, Mike and Davis, Mark and Kesselman, Joe and Pixley, Tom and Robie, Jonathan and Sharpe, Peter and Wilson, Chris},
  journal={World Wide Web Consortium. www. w3. org/TR/DOM-Level-2},
  year={2000}
}

@book{Lubanovic2023_FastAPI_misc,
  title={FastAPI},
  author={Lubanovic, Bill},
  year={2023},
  publisher={" O'Reilly Media, Inc."}
}

@incollection{Marcondes2025_Ollama,
  title={Using ollama},
  author={Marcondes, Francisco S and Gala, Adelino and Magalh{\~a}es, Renata and Perez de Britto, Fernando and Dur{\~a}es, Dalila and Novais, Paulo},
  booktitle={Natural Language Analytics with Generative Large-Language Models: A Practical Approach with Ollama and Open-Source LLMs},
  pages={23--35},
  year={2025},
  publisher={Springer}
}

@inproceedings{Xu2025_OnDeviceLLM,
  title={Fast on-device LLM inference with npus},
  author={Xu, Daliang and Zhang, Hao and Yang, Liming and Liu, Ruiqi and Huang, Gang and Xu, Mengwei and Liu, Xuanzhe},
  booktitle={Proceedings of the 30th ACM International Conference on Architectural Support for Programming Languages and Operating Systems, Volume 1},
  pages={445--462},
  year={2025}
}

@article{zou2023universal,
  title={Universal and transferable adversarial attacks on aligned language models},
  author={Zou, Andy and Wang, Zifan and Carlini, Nicholas and Nasr, Milad and Kolter, J Zico and Fredrikson, Matt},
  journal={arXiv preprint arXiv:2307.15043},
  year={2023}
}

@article{wei2023jailbroken,
  title={Jailbroken: How does llm safety training fail?},
  author={Wei, Alexander and Haghtalab, Nika and Steinhardt, Jacob},
  journal={Advances in Neural Information Processing Systems},
  volume={36},
  pages={80079--80110},
  year={2023}
}

@article{perez2022ignore,
  title={Ignore previous prompt: Attack techniques for language models},
  author={Perez, F{\'a}bio and Ribeiro, Ian},
  journal={arXiv preprint arXiv:2211.09527},
  year={2022}
}

@inproceedings{bagdasaryan2023instruction,
  title={Instructions as backdoors: Backdoor vulnerabilities of instruction tuning for large language models},
  author={Xu, Jiashu and Ma, Mingyu and Wang, Fei and Xiao, Chaowei and Chen, Muhao},
  booktitle={Proceedings of the 2024 Conference of the North American Chapter of the Association for Computational Linguistics: Human Language Technologies (Volume 1: Long Papers)},
  pages={3111--3126},
  year={2024}
}

@article{qi2023visual,
  title={Visual adversarial examples jailbreak large language models},
  author={Qi, Xiangyu and Huang, Kaixuan and Panda, Ashwinee and Wang, Mengdi and Mittal, Prateek},
  journal={CoRR},
  year={2023}
}

@article{bailey2023image,
  title={Image hijacks: Adversarial images can control generative models at runtime},
  author={Bailey, Luke and Ong, Euan and Russell, Stuart and Emmons, Scott},
  journal={arXiv preprint arXiv:2309.00236},
  year={2023}
}

@article{goodfellow2015explaining,
  title={Explaining and harnessing adversarial examples},
  author={Goodfellow, Ian J and Shlens, Jonathon and Szegedy, Christian},
  journal={arXiv preprint arXiv:1412.6572},
  year={2014}
}

@inproceedings{madry2018towards,
  title={Towards Deep Learning Models Resistant to Adversarial Attacks},
  author={Madry, Aleksander and Makelov, Aleksandar and Schmidt, Ludwig and Tsipras, Dimitris and Rogers, Adrian},
  booktitle={International Conference on Learning Representations},
  year={2018}
}

@inproceedings{cohen2019certified,
  title={Certified adversarial robustness via randomized smoothing},
  author={Cohen, Jeremy and Rosenfeld, Elan and Kolter, Zico},
  booktitle={international conference on machine learning},
  pages={1310--1320},
  year={2019},
  organization={PMLR}
}

@article{jain2023baseline,
  title={Baseline defenses for adversarial attacks against aligned language models},
  author={Jain, Neel and Schwarzschild, Avi and Wen, Yuxin and Somepalli, Gowthami and Kirchenbauer, John and Chiang, Ping-yeh and Goldblum, Micah and Saha, Aniruddha and Geiping, Jonas and Goldstein, Tom},
  journal={arXiv preprint arXiv:2309.00614},
  year={2023}
}

@article{alon2023detecting,
  title={Detecting language model attacks with perplexity},
  author={Alon, Gabriel and Kamfonas, Michael},
  journal={arXiv preprint arXiv:2308.14132},
  year={2023}
}

@book{zalewski2011tangled,
  title={The Tangled Web: A Guide to Securing Modern Web Applications},
  author={Zalewski, Michal},
  publisher={No Starch Press},
  year={2011}
}

@inproceedings{barth2008robust,
  title={Robust defenses for cross-site request forgery},
  author={Barth, Adam and Jackson, Collin and Mitchell, John C},
  booktitle={Proceedings of the 15th ACM conference on Computer and communications security},
  pages={75--88},
  year={2008}
}

@inproceedings{reis2019site,
  title={Site isolation: Process separation for web sites within the browser},
  author={Reis, Charles and Moshchuk, Alexander and Oskov, Nasko},
  booktitle={28th USENIX Security Symposium (USENIX Security 19)},
  pages={1661--1678},
  year={2019}
}

@inproceedings{lekies2013flows,
  title={25 million flows later: large-scale detection of DOM-based XSS},
  author={Lekies, Sebastian and Stock, Ben and Johns, Martin},
  booktitle={Proceedings of the 2013 ACM SIGSAC conference on Computer \& communications security},
  pages={1193--1204},
  year={2013}
}

@article{kang2017neurosurgeon,
  title={Neurosurgeon: Collaborative intelligence between the cloud and mobile edge},
  author={Kang, Yiping and Hauswald, Johann and Gao, Cao and Rovinski, Austin and Mudge, Trevor and Mars, Jason and Tang, Lingjia},
  journal={ACM SIGARCH Computer Architecture News},
  volume={45},
  number={1},
  pages={615--629},
  year={2017},
  publisher={ACM New York, NY, USA}
}

@inproceedings{teerapittayanon2017distributed,
  title={Distributed deep neural networks over the cloud, the edge and end devices},
  author={Teerapittayanon, Surat and McDanel, Bradley and Kung, Hsiang-Tsung},
  booktitle={2017 IEEE 37th international conference on distributed computing systems (ICDCS)},
  pages={328--339},
  year={2017},
  organization={IEEE}
}

@inproceedings{li2019edge,
  title={Edge intelligence: On-demand deep learning model co-inference with device-edge synergy},
  author={Li, En and Zhou, Zhi and Chen, Xu},
  booktitle={Proceedings of the 2018 workshop on mobile edge communications},
  pages={31--36},
  year={2018}
}

@inproceedings{laskaridis2020spinn,
  title={SPINN: Synergistic progressive inference of neural networks over device and cloud},
  author={Laskaridis, Stefanos and Venieris, Stylianos I and Almeida, Mario and Leontiadis, Ilias and Lane, Nicholas D},
  booktitle={Proceedings of the 26th annual international conference on mobile computing and networking},
  pages={1--15},
  year={2020}
}

@inproceedings{yao2023react,
  title={React: Synergizing reasoning and acting in language models},
  author={Yao, Shunyu and Zhao, Jeffrey and Yu, Dian and Du, Nan and Shafran, Izhak and Narasimhan, Karthik R and Cao, Yuan},
  booktitle={The eleventh international conference on learning representations},
  year={2022}
}

@article{schick2023toolformer,
  title={Toolformer: Language models can teach themselves to use tools},
  author={Schick, Timo and Dwivedi-Yu, Jane and Dess{\`\i}, Roberto and Raileanu, Roberta and Lomeli, Maria and Hambro, Eric and Zettlemoyer, Luke and Cancedda, Nicola and Scialom, Thomas},
  journal={Advances in Neural Information Processing Systems},
  volume={36},
  pages={68539--68551},
  year={2023}
}

@article{shen2023hugginggpt,
  title={Hugginggpt: Solving ai tasks with chatgpt and its friends in hugging face},
  author={Shen, Yongliang and Song, Kaitao and Tan, Xu and Li, Dongsheng and Lu, Weiming and Zhuang, Yueting},
  journal={Advances in Neural Information Processing Systems},
  volume={36},
  pages={38154--38180},
  year={2023}
}

@inproceedings{nvidia2023nemo,
  title={Nemo guardrails: A toolkit for controllable and safe llm applications with programmable rails},
  author={Rebedea, Traian and Dinu, Razvan and Sreedhar, Makesh Narsimhan and Parisien, Christopher and Cohen, Jonathan},
  booktitle={Proceedings of the 2023 conference on empirical methods in natural language processing: system demonstrations},
  pages={431--445},
  year={2023}
}

@inproceedings{iqbal2020adgraph,
  title={Adgraph: A graph-based approach to ad and tracker blocking},
  author={Iqbal, Umar and Snyder, Peter and Zhu, Shitong and Livshits, Benjamin and Qian, Zhiyun and Shafiq, Zubair},
  booktitle={2020 IEEE Symposium on security and privacy (SP)},
  pages={763--776},
  year={2020},
  organization={IEEE}
}

@misc{owasp2023llm,
  title={OWASP Top 10 for Large Language Model Applications},
  author={{OWASP}},
  year={2025},
  howpublished={\url{https://owasp.org/www-project-top-10-for-large-language-model-applications/}}
}

\end{document}